\documentclass[12pt]{article}
\usepackage{amsmath}
\usepackage{amssymb}
\usepackage{amsthm}
\usepackage{subcaption}
\usepackage[affil-it]{authblk}
\usepackage{natbib}
\usepackage[hyphens]{url}
\usepackage{hyperref}
\usepackage{graphicx}
\usepackage{multirow}
\usepackage{makecell}
\usepackage{threeparttable}
\usepackage{booktabs}
\usepackage{bm}
\usepackage{float}
\usepackage{pdfpages}
\usepackage{verbatim}
\usepackage{lineno}
\usepackage{enumitem}
\usepackage[english]{babel}
\usepackage{amsthm}
\usepackage{comment}

 \usepackage{color}

\newcommand{\bG}{\pmb{G}}

\newcommand{\var}{\mathrm{Var}}
\newcommand{\cov}{\mathrm{Cov}}
\newcommand{\cor}{\mathrm{Corr}}
\newcommand{\be}{{\epsilon}}
\newcommand{\bgamma}{\pmb{\gamma}}
\newcommand{\bbeta}{\pmb{\beta}}
\usepackage[top=1in, bottom=1in, left=0.96in, right=0.96in]{geometry}

\numberwithin{equation}{section}

\setcitestyle{super,open={},close={}}

\begin{document}
\author[*1]{Shuang Song}
\author[2,3]{Stefania Benonisdottir}
\author[4]{Jun S. Liu}
\author[2]{Augustine Kong \thanks{Correspondence: \texttt{shuangsong@hsph.harvard.edu, augustine.kong@bdi.ox.ac.uk}}}

\affil[1]{Department of Biostatistics, Harvard T. H. Chan School of Public Health, Harvard University, Boston, MA, USA.}
\affil[2]{Leverhulme Centre for Demographic Science, Nuffield Department of Population Health,
University of Oxford, Oxford, UK.}
\affil[3]{University of Iceland, Reykjavik, Iceland. }
\affil[4]{Department of Statistics, Harvard University, Cambridge, MA, USA.}

\date{}
\title{Participation bias in the estimation of heritability and genetic correlation}
\maketitle

\begin{abstract}
It is increasingly recognized that participation bias can pose problems for genetic studies. Recently, to overcome the challenge that genetic information of non-participants is unavailable, it is shown that by comparing the IBD (identity by descent) shared and not-shared segments between participating relative pairs, one can estimate the genetic component underlying participation. That, however, does not directly address how to adjust estimates of heritability and genetic correlation for phenotypes correlated with participation. Here, for phenotypes whose mean differences between population and sample are known, we demonstrate a way to do so by adopting a statistical framework that separates the genetic and non-genetic correlations between participation and these phenotypes. Crucially, our method avoids making the assumption that the effect of the genetic component underlying participation is manifested entirely through these other phenotypes. Applying the method to 12 UK Biobank phenotypes, we found 8 that have significant genetic correlations with participation, including body mass index, educational attainment, and smoking status. For most of these phenotypes, without adjustments, estimates of heritability and the absolute value of genetic correlation would have underestimation biases.
\end{abstract}


\section{Main}
The rapid development of biobank studies provides an unprecedented opportunity for understanding the genetics across many phenotypes \citep{sudlow2015uk}. At the same time,
 significant effort has been devoted to issues that complicate data analyses. These include  population stratification, cryptic relatedness, assortative mating, and measurement error \citep{bulik2015ld, kong2018nature, abdellaoui2021dissecting, border2022assortative}. 
 More recently, particularly for investigations that go beyond the testing of associations between phenotypes and individual genetic variants, it is increasingly appreciated that  participation bias (PB), $i.e.$ the ascertained samples are not fully representative of the population, can lead to misleading results \citep{winship1992models}. While PB is a concern for all sampling surveys, it is particularly difficult to avoid for genetic studies given the requirement of informed consent and the collection of DNA material. For example, despite the large sample size of UK Biobank (UKBB), the participation rate of those who were invited is only about 
 $5.5\%$ \citep{swanson2012uk}, allowing for the possibility of substantial bias in some aspects of the data.

For sample surveys in general, a common approach to adjusting for ascertainment bias is to construct a propensity score based on variables, \emph{e.g.} sex and educational attainment (EA), whose distributional differences between sample and target population are known, or can be well approximated \citep{seaman2013review}. By assuming that the systematic component of participation probability is fully captured by this score, analyses can be adjusted by applying inverse propensity weighting (IPW) to the samples. For genetic studies, it has been shown that participation can be associated with many phenotypes, such as EA, alcohol use, mental and physical health \citep{van2022reweighting, bisgard1994mortality, manjer2001malmo, drivsholm2006representativeness,  fry2017comparison, knudsen2010health, schoeler2022correction}. The UKBB participants are reported to be
less likely to be obese, to smoke, to drink alcohol daily, and to have fewer self-reported health conditions compared with  the general UK population \citep{fry2017comparison}. 
A recent study applied IPW to the UKBB data by creating a propensity score based on 14 phenotypes including age, BMI, weight, education, etc. \citep{schoeler2022correction}
This propensity score, however, does not include genotypes. Thus, for analyses that involve genotypes, this adjustment would be sufficient only under the assumption that the systematic genotypic differences between sample and population are fully captured by this propensity score. This is equivalent to saying that the effect of the genetic component underlying participation is manifested entirely through this score, which can be considered as a composite phenotype. Previous results\citep{benonisdottir2022genetics} and results presented below show that this assumption does not hold.


While the effectiveness of IPW or any other adjustment methods based only on phenotypes is doubtful, there is no obvious alternative unless genotype difference between sample and population can be independently estimated without relying on the phenotypes, which is difficult given that genotypes of non-participants are unavailable. A recent publication showed that the sample-population allelic frequency differences can be estimated by comparing the IBD (identity by descent) shared and not-shared segments among the participants \citep{benonisdottir2022genetics}. Here we utilize information obtained from that method through introducing a statistical model that specifies a genetic and a non-genetic component for each of the participation variable and each of the other phenotypes. The model separates the genetic and non-genetic correlations between participation and other correlated phenotypes, and allows us to obtain adjusted estimates of heritability and genetic correlation for these phenotypes that take PB into account. Biases of the estimated genetic components of the phenotypes and possible adjustments will also be discussed.

Theoretically, without adjustments, heritability and genetic correlation can be over-estimated or under-estimated depending on the relative magnitudes of the genetic and non-genetic correlations between participation and the phenotypes. Empirically, we applied the adjustment method to 12 phenotypes of the UKBB data and found that, without adjustment, heritability and the absolute value of the genetic correlation estimates tend to be underestimated for most of them.

\section{Results}
\subsection{Participation model overview}
In practice,  participation is often a two-step process. In the first step, a group of  individuals are invited to participate in the study, and in the second step the invited make the decision whether to participate. For simplicity, we consider a model where the bias is only in the second step, $i.e.$ the invited list is representative of the target population.
Consequences of the violation of this assumption are discussed later.
For the second step, as in Benonisdottir and Kong (2023)\citep{benonisdottir2022genetics}, we adopt a liability-threshold model where the liability score of a person is denoted by  $X$, which is assumed to have (approximately) a standard normal distribution in the population. An invited person participates if $X> t_\alpha$, where $t_\alpha=\Phi^{-1}(1-\alpha)$,  $\Phi$ is the  standard normal  cumulative distribution function  and $\alpha$ is the participation rate of those invited. A phenotype of interest is denoted by $Y$, standardized to have mean zero and variance 1. For an individual in the invited list, we assume it is a random draw following the following model:
\begin{equation}
\begin{aligned}
  X= G_x+ \be_x, \\
  Y= G_y+ \be_y,
  \end{aligned}
  \label{eq:model}
\end{equation}
where $G_x$ and $G_y$ are the  genetic components of $X$ and $Y$, respectively.  We assume that  they can be represented by  weighted sums of genotypes.
Here $\be_x$ and $\be_y$ are parts of $X$ and $Y$ that are orthogonal (uncorrelated) to the genetic components $G_x$ and $G_y$, which could be partly random and partly determined by non-genetic factors. 
We denote the heritability of $X$ and $Y$ in the population as $h^2_x$ and $h^2_y$, which measure the phenotypic variance explained by  genetic components. 
In model \eqref{eq:model}, we have $\var(G_x)=h^2_x$,  $\var(G_y)=h^2_y$, and  $\var(\be_x)=1-h^2_x$,  $\var(\be_y)=1-h^2_y$. 
The genetic and non-genetic correlations between participation liability score $X$ and phenotype $Y$ in the population are denoted as 
$\rho_g=\cor(G_x, G_y)$ and $\rho_e=\cor(\be_x,\be_y)$, respectively. The correlation between $X$ and $Y$, denoted by $\rho$, is equal to $\rho=\rho_g\sqrt{h^2_xh^2_y}+\rho_e\sqrt{(1-h^2_x)(1-h^2_y)}$.

\begin{figure}[H]
 \centering
 \includegraphics[width=0.8\textwidth]{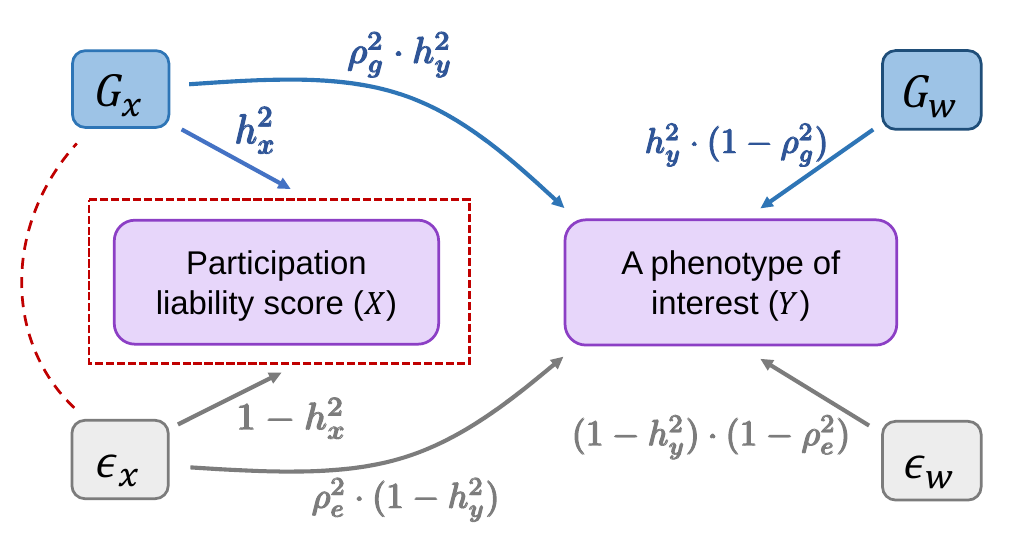}
  \caption{\textbf{Graphical representation of the contributions of genetic and non-genetic variables underlying the participation liability score $X$ and another phenotype $Y$.} The genetic and non-genetic components of $Y$ are reformulated as $G_y=a\cdot G_x+G_w$ and $\be_y=b\cdot \be_x+\be_w$. Variables $G_x, G_w, \be_x$, and $\be_w$ are orthogonal in the population. Each arrow is accompanied by the causal strength of the explanatory variable, quantified as the $r^2$ between the two variables. As $X$ and $Y$ are standardized, the contributions of the two arrows pointing to $X$ and the four arrows pointing to $Y$ each sum to 1. Upon conditioning on participation, $X>t_\alpha$, variances of $G_x$ and $\be_x$ are attenuated and they become negatively correlated (the red dashed line) as a result of collider bias.
 }\label{fig:graphical}
\end{figure}

To better illustrate the effect of PB, we reparametrize $G_y$ as $a\cdot G_x+G_w$, where $a=\rho_g\sqrt{h^2_y/h^2_x}$ and $G_w$ is orthogonal to $G_x$ with $\var(G_w)=(1-\rho^2_g)h^2_y$. Similarly, $\be_y$ is reformulated as $b\cdot \be_x+\be_w$, where $b=\rho_e\sqrt{(1-h^2_y)/(1-h^2_x)}$ and $\be_w$ is orthogonal to $\be_x$ with $\var(\be_w)=(1-\rho^2_e)(1-h^2_y)$. Figure \ref{fig:graphical} is a directed graph illustrating the mathematical causal relationships of the variables. The four explanatory variables $G_x$, $G_w$, $\be_x$, and $\be_w$ are independent of each other in the population. PB as a result of conditioning on $X>t_\alpha$, has the following impact on the sample. As $G_w$ and $\be_w$ are not causal variables of $X$, their distributions are not affected by the conditioning, and they remain uncorrelated with each other and with $G_x$ and $\be_x$. By contrast, the variances of $G_x$ and $\be_x$ shrink, and furthermore they become negatively correlated because of collider bias.

\subsection{Heritability estimated with participation bias}
\label{sec:heri}
Under model \eqref{eq:model}, the heritability of phenotype $Y$ can be calculated as $h^2_y=\cor(Y,G_x)^2$.
Let 
\begin{equation}
h^2_{y,PB}=\cor(Y, G'_y\mid X> t_\alpha)^2, 
\label{eq:h2.pb.def1}
\end{equation}
which is the ``apparent" heritability that is estimated using
the sample of participants without considering PB. 
Here $G'_y$ is the genetic component in the sample of participants, 
and $h^2_{y,PB}$ can be regarded as the expected proportion of the phenotypic variance that can be explained by genetic variants alone in the sample. Specifically, $G'_y$ can be written as $a'\cdot G_x+G_w$ (expression for $a'$ is given in the Supplementary Note 1.2). Consider here the more common case where $\rho_g$ and $\rho_e$ have the same sign. Suppose both $\rho_g$ and $\rho_e$ are positive, which implies $a$ and $b$ are positive (for cases where these parameters are all negative, the discussion here would apply to $-Y$). In this case, $a'<a$ because $G_x$ and $\be_x$ are negatively correlated upon conditioning, and thus the coefficient of $G_x$ is reduced because it is capturing some of the positive contribution of $\be_x$ to $Y$ in a negative manner. Since $a'\cdot G_x+G_w=G_y-(a-a')\cdot G_x$, this means that the genetic component of $Y$ estimated from the sample is biased towards reducing the contribution of the $G_x$ component. After selection, the variance of $Y$ explained by $G_w$ remains unchanged, while, ignoring rare cases where $a'<0$ and $|a'|>a$, the variance accountable by $G_x$ would decrease as both its variance and its coefficient are reduced. The variance of $Y$ also shrinks because of reduced contributions from both $G_x$ and $\be_x$. As both the variance accountable by genetic components and the total variance shrink, their ratio ($i.e.$ $h^2_{y,PB}$) can either increase or decrease depending on the relative shrinkages.

The quantitative results presented below are derived assuming that the genetic and non-genetic components ($G_x, G_w, \be_x, \be_w$) jointly follow a multivariate normal distribution (MVN), which should be appropriate for complex traits. The expressions appear more complicated than usual with MVN manipulations because conditioning is on an interval, instead of on a single value, of $X$. Based on the properties of truncated MVN\citep{birnbaum1950effect, kan2017moments}, we show that (see Supplementary Note 1.2)
\begin{equation}
h^2_{y,PB}=\frac{1}{1-\xi(\alpha)\rho^2}\left[h^2_y-
\xi(\alpha)\rho_G(\rho_G+2\rho_E)+\frac{\xi(\alpha)^2h^2_x\rho_E^2}{1-\xi(\alpha) h^2_x}\right],
  \label{eq:h2.pb}
\end{equation}
where $\rho_G=\rho_g\sqrt{h^2_xh^2_y}$ is the covariance of $G_x$ and $G_y$; $\rho_E=\rho_e\sqrt{(1-h^2_x)(1-h^2_y)}$ is the covariance of $\be_x$ and $\be_y$; and 
\begin{equation}
\label{eqn:xi}
\xi(\alpha)=\left[\frac{\phi(t_\alpha)}{\alpha}\right]^2-t_\alpha\cdot\left[\frac{\phi(t_\alpha)}{\alpha}\right],
\end{equation} 
where $\phi$ is the  density function of the standard normal distribution. Notably, $\left[1-\xi(\alpha)\right]=\var(X|X>t_\alpha)$ and the denominator of \eqref{eq:h2.pb}, $(1-\xi(\alpha)\rho^2)$, is the variance of $Y$ after selection. The numerator of \eqref{eq:h2.pb} is the variance of $Y$ that can be accounted for by $G_x$ and $G_w$ after selection.

\begin{figure}[H]
 \centering
 \includegraphics[width=\textwidth]{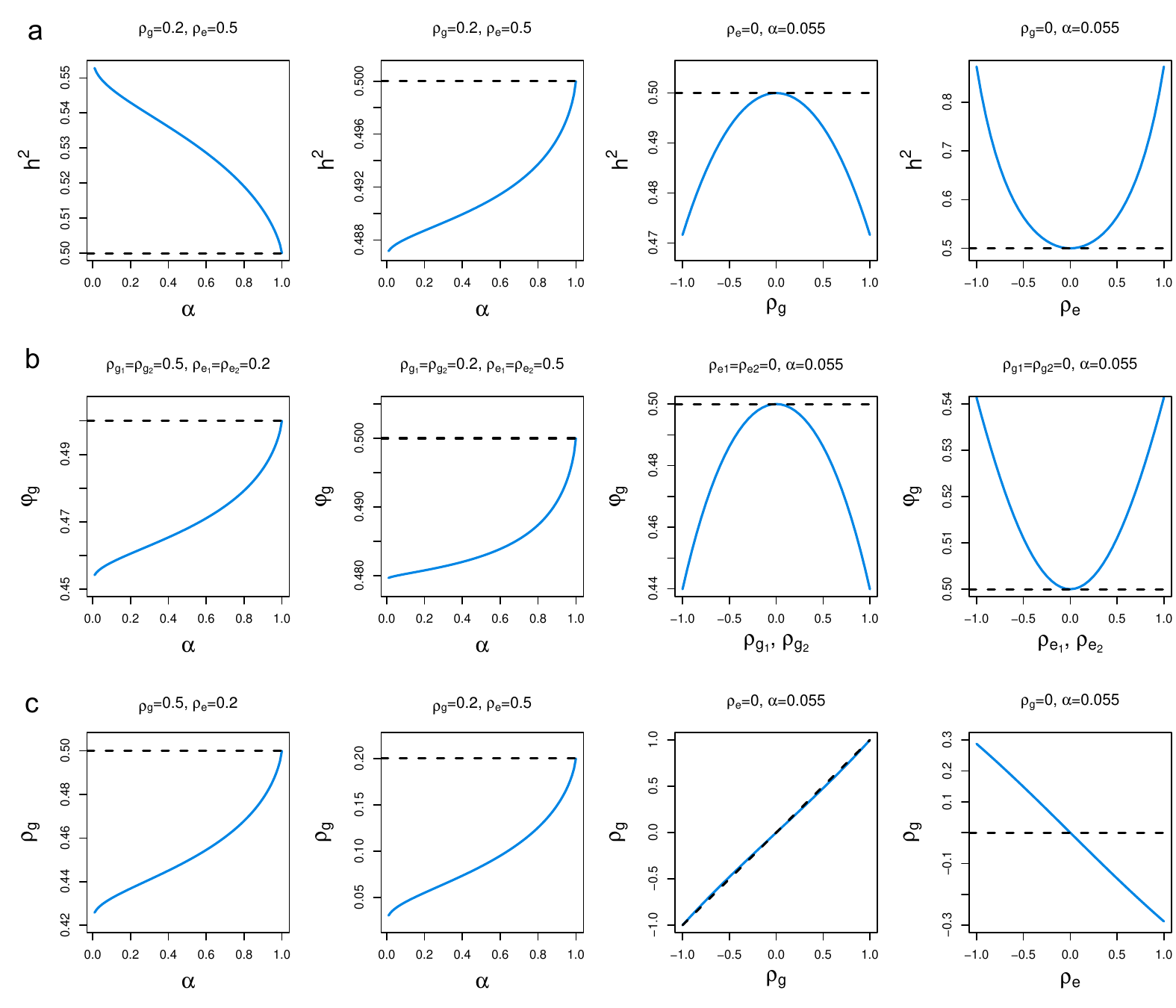}
  \caption{\textbf{Theoretical results for heritability and genetic correlation in the sample under various genetic architectures.} 
   (a) The heritability of a phenotype of interest other than participation. (b) The genetic correlation between two phenotypes (other than participation). (c) The genetic correlation between the liability score of participation and the phenotype.
 The heritability of participation liability score and other phenotypes are fixed at 0.125 and 0.5. Dashed lines indicate population value and blue solid lines indicate what is being estimated based on the biased sample without adjustment.
 }\label{fig:h2.est}
\end{figure}

A comparison of $h^2_y$ and $h^2_{y,PB}$ is shown in 
Figure \ref{fig:h2.est}(a).
In general, with  other parameters fixed, the difference between $h^2_y$ and $h^2_{y,PB}$ increases as $\alpha$ decreases. As noted earlier, PB can lead to both upward and downward biases for the estimation of heritability in the  sample of participants. For example, if the correlation between $X$ and $Y$ arises solely from genetic components, conditioning on participation will reduce the variance of the genetic component of $Y$  in the selected sample.
With the variance of the non-genetic component  unchanged, 
the proportion of the genetic effects in the  sample will shrink. Conversely, if the correlation between the phenotype and participation is induced by non-genetic factors only, the proportion of variance of $Y$ attributed to the genetic component in the sample will increase, which leads to an upward bias in the estimation of heritability.

\subsection{Genetic correlation estimated with participation bias}
\label{sec:gcov}
Here we consider
 two phenotypes, $Y_1$ and $Y_2$, both following the model \eqref{eq:model}, and  $\varphi_g$ denotes their genetic correlation in the population.
 The corresponding genetic covariance is  $\varphi_G=\varphi_g\sqrt{h^2_{y_1}h^2_{y_2}}$, where $h^2_{y_1}$ and $h^2_{y_2}$ are the heritability of the two phenotypes.
Let $\varphi_{g,PB}=\cor(G'_{y_1}, G'_{y_2}| X > t_\alpha)$.
In parallel with  $h^2_{y,PB}$, the quantity $\varphi_{g,PB}$ represents the ``apparent"  correlation between the genetic components of $Y_1$ and $Y_2$ that is being estimated ignoring PB. 
We show in Supplementary Note 1.3 that,
\begin{equation}
  \varphi_{g,PB} = \frac{\varphi_{G}-\xi(\alpha)(\rho_{E_1}\rho_{G_2}+\rho_{E_2}\rho_{G_1}+\rho_{G_1}\rho_{G_2})+
\frac{\xi(\alpha)^2h^2_x\rho_{E_1}\rho_{E_2}}{1-\xi(\alpha) h^2_x}}{\sqrt{(1-\xi(\alpha)\rho^2_{1})(1-\xi(\alpha)\rho^2_{2})}\cdot \sqrt{h^2_{y_1,PB}h^2_{y_2,PB}}}.
\label{eq:phi_gPB}
\end{equation}
Here $\rho_{G_1}$ and $\rho_{G_2}$ are the  covariance of $G_x$ with  $G_{y_1}$ and $G_{y_2}$; 
$\rho_{E_1}$ and $\rho_{E_2}$ are the  covariance of $\be_x$ with  $\be_{y_1}$ and $\be_{y_2}$; 
$\rho_{1}$ and $\rho_{2}$ are the phenotypic correlations between $X$ and $Y_1$ and $Y_2$, respectively, in the population. 
The denominator is the squared root of the product of variances explained by $G'_{y_1}$ and $G'_{y_2}$ in the sample of participants, where the heritabilities $h^2_{y_1,PB}$ and $h^2_{y_2,PB}$ are derived in \eqref{eq:h2.pb}.
The numerator is the  covariance between $G'_{y_1}$ and $G'_{y_2}$, which incorporates the negative correlations between $G_{x_1}$ and $\be_{x_1}$, and  $G_{x_2}$ and $\be_{x_2}$ in the sample of participants.  

Figure \ref{fig:h2.est}(b) shows the comparison of $\varphi_g$ and $\varphi_{g,PB}$. 
Similar to the results of heritability estimates, the bias induced by PB on genetic correlation increases when the participation rate decreases.
The estimation of genetic correlation can have either upward or downward bias in the sample of participants.

\subsection{Genetic correlation between  participation and a phenotype}
\label{sec:gcov_part}
Intuitively, both genetic and non-genetic correlations between participation and the phenotype contribute to the overall correlation, which induces PB when selection occurs based on participation liability score.
We define $\rho_{g,PB}=\cor(G_x,G'_y|X> t_\alpha)$, where $G'_y=a'\cdot G_x+G_w$ is defined in Section \ref{sec:heri}.
We note that $\rho_{g,PB}$ is a special case of $\varphi_{g,PB}$, by regarding $Y_1$ as $X$, and $Y_2$ as $Y$. Specifically, we have $h^2_{y_1}=h^2_x$, $h^2_{y_2}=h^2_y$, $\rho_1=\rho_{g_1}=\rho_{e_1}=1$,  $\rho_2=\rho$, $\varphi_g=\rho_{g_2}=\rho_g$, and $\varphi_e=\rho_{e_2}=\rho_e$.
It can be shown that (see Supplementary Note 1.4)
\begin{equation}
  \rho_{g,PB} = \frac{\rho_G-\xi(\alpha)h^2_x\rho }{\sqrt{(1-\xi(\alpha)\rho^2)(1-\xi(\alpha)h^2_x)}\cdot\sqrt{h^2_xh^2_{y,PB}}},
  \label{eq:rho_g_prime}
\end{equation}
where the denominator is the squared root of the product of the variance explained by $G_x$ and $G'_y$ in the sample of participants, and the numerator is their covariance. When $X$ and $Y$ are positively correlated (i.e. $\rho>0$), the numerator  is smaller than $\rho_G$ due to the reduced proportion of $G_x$ in $G'_y$ after selection. 
Some comparisons between $\rho_{g,PB}$ and $\rho_g$ are in  Figure \ref{fig:h2.est}(c). 
Note that if $\rho_g=0$, $\rho_{g,PB}$ will be in the opposite direction of $\rho$ (or $\rho_e$) because of collider bias.

\subsection{Estimating heritability and genetic correlation  with PB adjustment}
\label{sec:res.correct}
With the results derived above, we provide a method to correct the effects of PB on the estimates of heritability and genetic correlations. In addition to the standard estimates computed using the sample, our method utilizes two pieces of information: \romannumeral1) the genetic test statistics for participation derived from the IBD-based comparisons \citep{benonisdottir2022genetics}; and \romannumeral2)
the mean shifts of other phenotypes from population to sample.
Notably, although genotypes of non-participants are unavailable, population average of many phenotypes are available from sources such as census data. We show below how to adjust the estimate of genetic covariance and heritability. 

We define the mean shift of a standardized phenotype between the selected sample and the population 
as $\delta=\left[\mathbb{E}(Y|X> t_\alpha)-\mathbb{E}(Y)\right]/\sqrt{\var(Y|X>t_\alpha)}$. While most mathematics have been derived for variables standardized with respect to the population,   here the mean shift is standardized with respect to the sample for the convenience of applications. For model \eqref{eq:model},
\begin{equation}
\delta=\frac{\rho\phi\left(t_\alpha\right)}{\alpha\sqrt{1-\xi(\alpha)\rho^2}}.
\label{eq:delta}
\end{equation}
Given $\alpha$, $\delta$ is a monotonically increasing function of $\rho$ (Supplementary Figure 1), 
and given $\rho$, $\delta$  increases as $\alpha$ decreases.  

In practice, we estimate the mean phenotypic value in both the sample of participants and  another cohort that is representative of the population. The observed mean shift $\widehat{\delta}$ is just  the difference of the two estimates standardized with respect to the sample of participants. 
Based on \eqref{eq:delta}, we derive the estimate of the phenotypic correlation between $X$ and $Y$:
\begin{equation}
\widehat{\rho}=\frac{\alpha\widehat{\delta}}{\sqrt{\xi(\alpha) \alpha^2\widehat{\delta}^2+\phi\left(t_\alpha\right)^2}}.
\label{eq:rho_hat}
\end{equation}
For notations,   we use 
$\widehat{h}^2_y$,
$\widehat{\rho}_g$, and$\widehat{\varphi}_g$ to denote the estimates that are affected by but not adjusted for PB.
We  denote the corresponding  adjusted estimates  as 
$\widetilde{h}^2_y$, $\widetilde{\rho}_g$, 
and $\widetilde{\varphi}_g$. 
By solving $\rho_g$ from (20)  of Supplementary Note 1.4, we derive the adjusted genetic covariance of $X$ and $Y$:
\begin{equation}
\label{eq:correct.gcov}
  \widetilde{\rho}_G=\sqrt{1-\xi(\alpha)\widehat{\rho}^2}\cdot\widehat{\rho}_{G}+\xi(\alpha)\widehat{\rho}\widehat{h}^2_{x},
\end{equation}
where $\widehat{h}^2_{x}$ is the SNP heritability of the participation liability score estimated with IBD-based information \citep{benonisdottir2022genetics}. The genetic covariance between participation liability score and the phenotype can be derived from (21) of Supplementary Note 1.4:
\begin{equation}
\widehat{\rho}_G=\widehat{\rho}_g\sqrt{(1-\xi(\alpha)\widehat{h}^2_{x})\cdot \widehat{h}^2_{x}\cdot \widehat{h}^2_{y}}.
\label{eq:rho_G_hat}
\end{equation}
By solving $h^2_y$ from \eqref{eq:h2.pb} and substituting parameters with their estimates,  the adjusted heritability estimate of $Y$ is:
\begin{equation}
  \widetilde{h}^2_y=\widehat{h}^2_y(1-\xi(\alpha)\widehat{\rho}^2)+2\xi(\alpha)\widehat{\rho}\widetilde{\rho}_G-\xi(\alpha)^2\widehat{\rho}^2\widehat{h}^2_{x}-\frac{\xi(\alpha)}{1-\xi(\alpha)\widehat{h}^2_{x}}\left(\widetilde{\rho}_G-\xi(\alpha)\widehat{\rho}\widehat{h}^2_{x}\right)^2.
  \label{eq:h2_correct}
\end{equation}
The adjusted genetic correlation of $X$ and $Y$ is:
\begin{equation}
\label{eq:correct.gcor}
  \widetilde{\rho}_g=\frac{\widetilde{\rho}_G}{\sqrt{\widehat{h}^2_{x}\widetilde{h}^2_y}}.
\end{equation}

Similarly, we solve for $\varphi_G$ from the approximation of $\varphi_{G,PB}$ in (18) in Supplementary Note 1.3, and substitute parameters with their estimates, leading to:
\begin{equation}
\begin{aligned}
  \widetilde{\varphi}_G=&\sqrt{(1-\xi(\alpha)\widehat{\rho}^2_{1})(1-\xi(\alpha)\widehat{\rho}^2_{2})}\widehat{\varphi}_{G}+\xi(\alpha)(\widehat{\rho}_{1}\widehat{\rho}_{G_2}+\widehat{\rho}_{2}\widehat{\rho}_{G_1})-\xi(\alpha)^2\widehat{\rho}_{1}\widehat{\rho}_{2}\widehat{h}^2_{x}\\
  &-\frac{\xi(\alpha)}{1-\xi(\alpha)\widehat{h}^2_{x}}(\widehat{\rho}_{G_1}-\xi(\alpha)\widehat{\rho}_{1}\widehat{h}^2_{x})(\widehat{\rho}_{G_2}-\xi(\alpha)\widehat{\rho}_{2}\widehat{h}^2_{x}),
\end{aligned}
\end{equation}
where $\widehat{\varphi}_{G}=\widehat{\varphi}_{g}\sqrt{\widehat{h}^2_{y_1}\widehat{h}^2_{y_2}}$ is the estimated genetic covariance of $Y_1$ and $Y_2$ ignoring PB; $\widehat{\rho}_{G_1}$ and $\widehat{\rho}_{G_2}$ are the estimated genetic covariance of $X$ and $Y_1$ and $Y_2$ ignoring PB, which can be derived from \eqref{eq:rho_G_hat}. The $\widehat{\rho}_{1}$ and $\widehat{\rho}_{2}$ are the estimated phenotypic correlation with participation for the two phenotypes derived with \eqref{eq:rho_hat}. The adjusted genetic correlation estimate is then:
\begin{equation}
    \widetilde{\varphi}_g=\frac{\widetilde{\varphi}_G}{\sqrt{\widetilde{h}^2_{y_1}\widetilde{h}^2_{y_2}}},
\end{equation}
with the denominator computed with \eqref{eq:h2_correct}.

The standard errors are estimated with a block jackknife procedure. 
In theory, if $\delta$ and $h^2_x$ are estimated with small standard errors, the adjustment could even lead to a reduced variance, $i.e.$ giving an estimate with both reduced bias and smaller standard error.

In practice, the heritability and genetic correlation are often estimated with marker-based methods such as LD score regression (LDSC). With random sampling, LDSC provides unbiased heritability and genetic covariance estimates \citep{bulik2015ld, ni2018estimation}, under the assumption that SNPs that are physically far apart are in linkage equilibrium (uncorrelated). In the selected sample, however, not only do the SNP effects change due to PB, but also the LD structure. In particular, SNPs with positive effects on the participation liability score $X$ that are uncorrelated in the population would become negatively correlated in the sample. While this change of LD does not affect \eqref{eq:h2.pb}, \eqref{eq:phi_gPB}, and \eqref{eq:rho_g_prime}, among other things, the LDSC estimate of  $h^2_{y,PB}$ would be negatively biased when $G_x$ and $G_y$ are correlated (see Supplementary Note 1.5). 
Note that this bias is mathematically similar to the positive bias induced by assortative mating \cite{border2022assortative}, only that they are in opposite direction. There are ways to further adjust for this bias if deemed necessary (see Supplementary Note 1.5), but we decided not to do so for the UKBB analyses that follow for two reasons. Firstly, in simulations where the population data are generated under the LDSC assumptions, we found that LDSC estimates of the sample parameters, $e.g.$ $h^2_{y,PB}$, are only very slightly biased relative to \eqref{eq:h2.pb}, \eqref{eq:phi_gPB}, and \eqref{eq:rho_g_prime}. Furthermore, the adjusted estimates, $\widetilde{h^2_y}$, $\widetilde{\rho_g}$, and $\widetilde{\varphi_g}$, calculated using the LDSC sample estimates are also only very slightly biased relative to the population parameters  (see  Supplementary Note 1.7 and Supplementary Table 1). Secondly, we examined the odd-chromosome and even-chromosome components of the polygenic score for $X$ and detected neither positive nor negative correlation $p>0.05$ with the UKBB data \citep{benonisdottir2022genetics} . Thus, given that the genetic components of $X$ and EA are positively correlated \citep{benonisdottir2022genetics}, it is plausible that this is a case where some assortative mating effect of $X$ partially cancels out the PB effect on distant LD.


\subsection{Analyses of UKBB data}
We applied our adjustment method to 12 UKBB phenotypes including 4 physical measures:
body mass index (BMI), height (HGT), waist circumference (WC), and hip circumference (HC);
3 sociodemographic measures:
EA, employment status (ES), and income (INC); and 5 lifestyle measures:
current smoking status (SMC), previous smoking status (SMP), alcohol consumption (ALC), walking pace (WKP), and walking time (WKT). Detailed description of the data is summarized in Supplementary Note 1.6. We first estimated heritability and genetic correlations across the phenotypes ignoring PB.
We also estimated their genetic correlation with participation based on the genome-wide association studies (GWAS) summary statistics derived from IBD-based information\cite{benonisdottir2022genetics}.
Then we calculated the mean shift between the UKBB and the Health Survey for England (HSE) dataset (Methods). We use the HSE dataset as a baseline as it incorporated the weighting to account for nonresponse bias.

\begin{figure}[H]
 \centering
 \includegraphics[width=\textwidth]{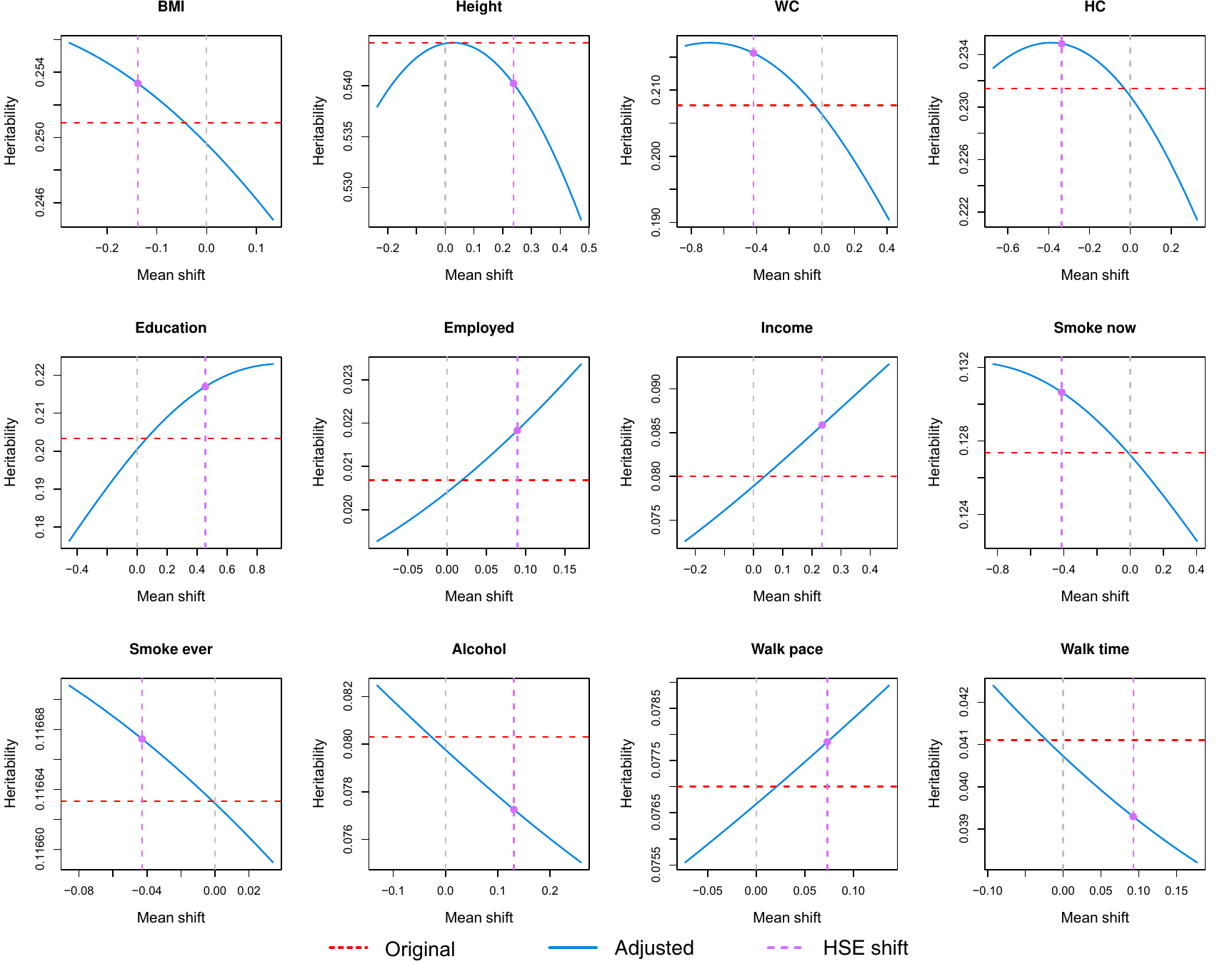}
  \caption{\textbf{Adjusted heritability estimates of 12 UKBB phenotypes.} The red dashed line indicates the original heritability estimates. The blue solid line indicates the heritability estimate adjusted for PB across varying mean shifts. The purple dashed line marks the mean shift between UKBB and HSE.
 }\label{fig:ukb.h2.curve}
\end{figure}

\begin{figure}[H]
 \centering
 \includegraphics[width=\textwidth]{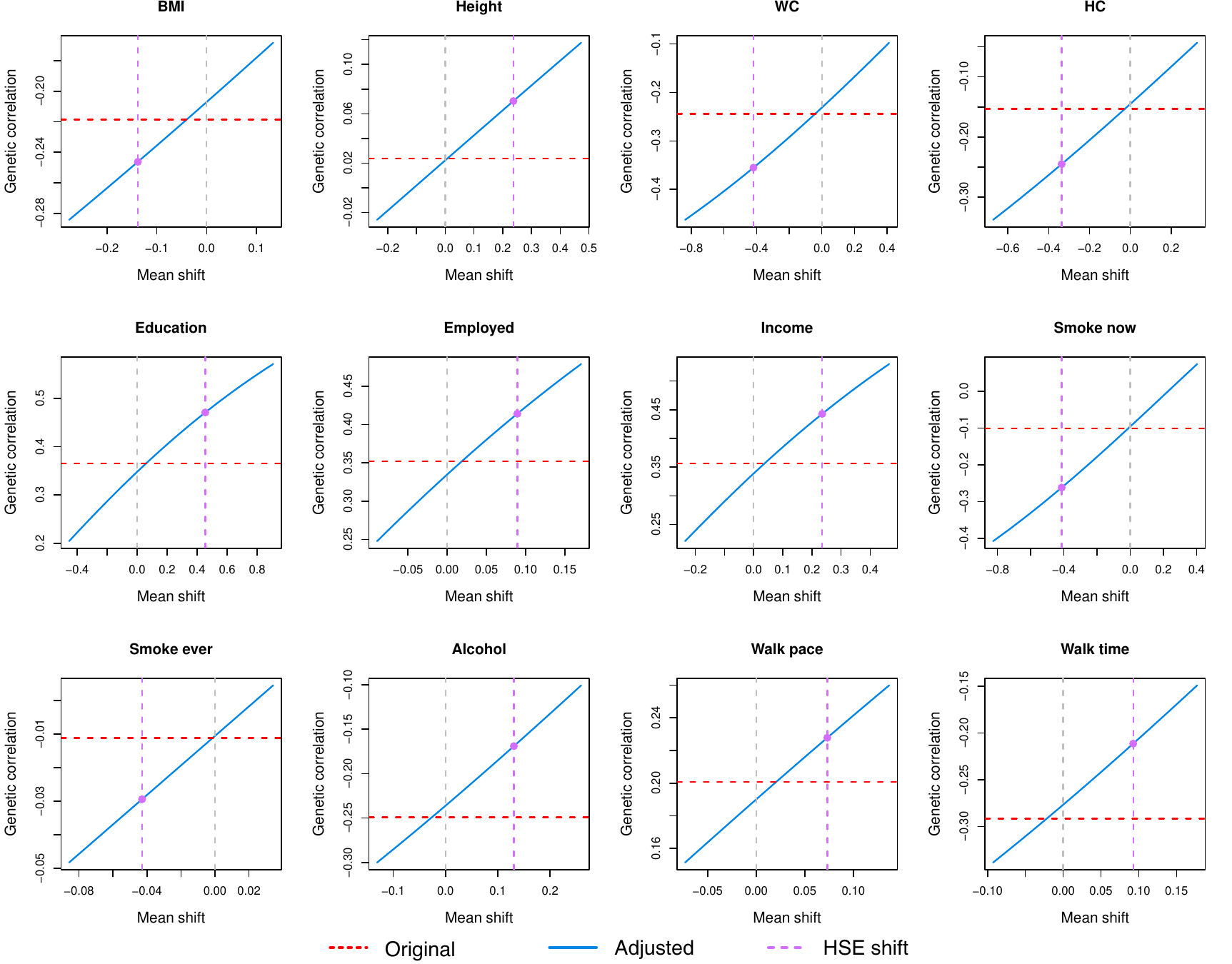}
  \caption{\textbf{Adjusted genetic correlation estimates between participation and the 12 UKBB phenotypes.} The red dashed line indicates the original genetic correlation estimates. The blue solid line indicates the genetic correlation estimate adjusted for PB across varying mean shifts. The purple dashed line marks the mean shift between UKBB and HSE. 
 }\label{fig:ukb.gcor.curve}
\end{figure}

Figures \ref{fig:ukb.h2.curve} and \ref{fig:ukb.gcor.curve} show the effect of PB on the estimates of heritability and genetic correlation of the UKBB phenotypes as functions of mean shifts of the phenotypes. The adjusted estimates based on the observed mean shifts between UKBB and HSE data are indicated.  
Numerical results of the original (unadjusted) and adjusted estimates are shown in Table \ref{tab:realdata}. 
With adjustments, the genetic components of BMI, WC, HC, EA, ES, INC, SMC, and WKP are significantly correlated with the genetic components of participation. Specifically, we found the heritability estimates have underestimation bias across those phenotypes that are genetically correlated with participation. Underestimation bias for unadjusted estimates is also observed for the absolute values of genetic correlations. In addition, 
 we found that SMC, which previously lacked significant genetic correlation with participation ($p=0.235$, two-sided test), surpassed the significance threshold of 0.05 with adjustment ($p=0.002$, two-sided test). For ALC and WKT, the unadjusted results showed  significant genetic correlation with participation ($p=0.006$ and $0.014$, two-sided test). However,  with adjustments, the estimated correlation shrunk and no longer statistically significant  ($p=0.104$ and $0.113$, two-sided test). None of the estimated correlations switched signs with adjustment for those phenotypes. 

\begin{figure}[H]
 \centering
 \includegraphics[width=\textwidth]{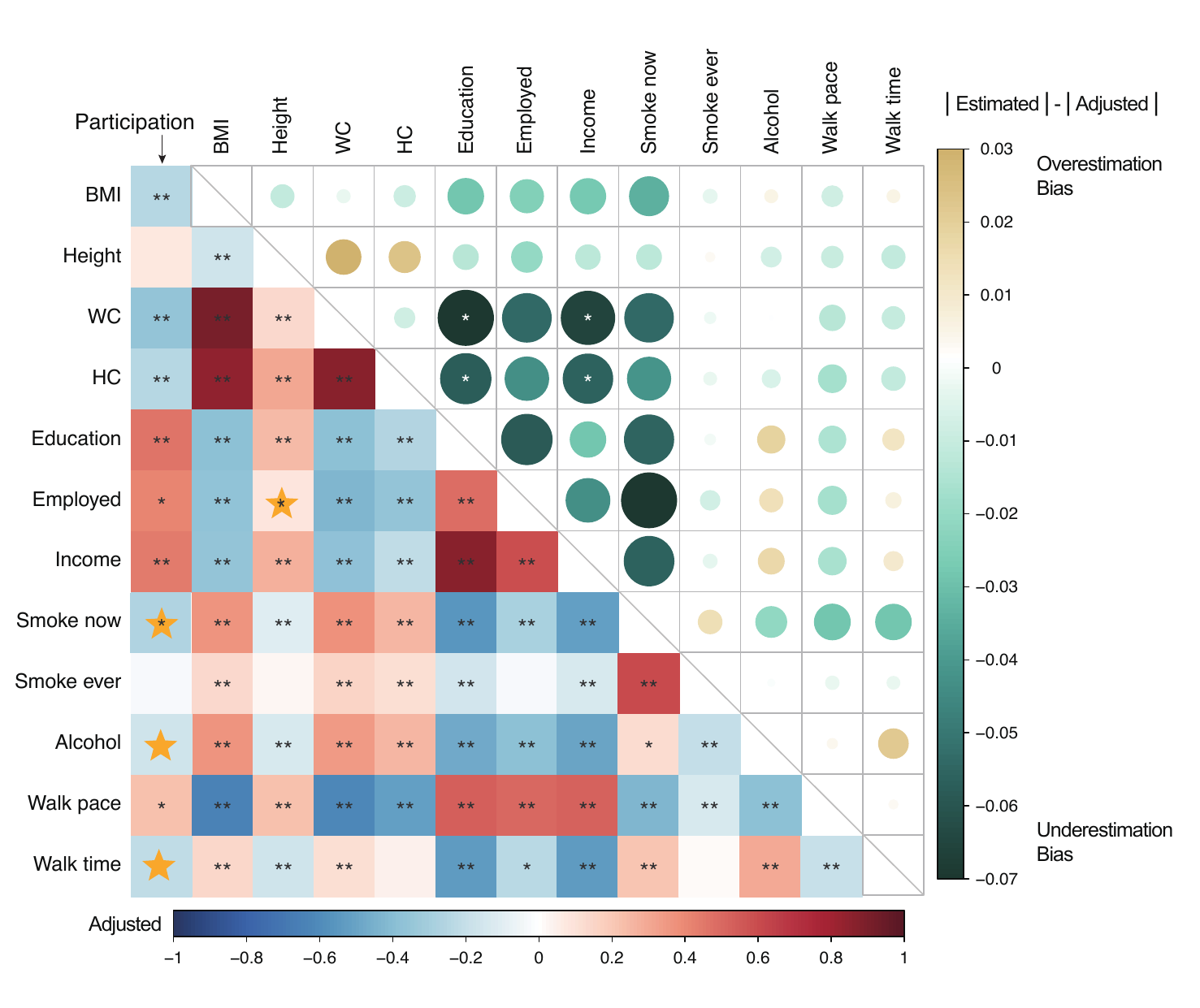}
  \caption{\textbf{The estimates of genetic correlations across 12 UKBB phenotypes before and after adjustment for PB.} The lower left panel: The adjusted genetic correlations. The asterisks highlight the significance level of genetic correlation after adjustment ($*$: $p<0.05$; $**$: $p<0.05/78$ (Bonferroni correction)). Yellow stars highlight cases where significance changed after adjustment, either from significant (threshold of 
$p<0.05$) to non-significant, or vice versa. The upper right panel:   the difference between the absolute value of the genetic correlation before and after adjustment. The size of each circle corresponds to the magnitude of the absolute difference, with larger circles indicating greater differences.  The asterisks highlight the significance level of the difference ($*$: $p<0.05$). 
 }\label{fig:gcor.realdata}
\end{figure}

We further analyzed the genetic correlation of each pair of the 12 phenotypes (Figure \ref{fig:gcor.realdata}). 
No estimate switched signs with adjustment. For phenotypes 
significantly correlated with participation, absolute values of all the unadjusted genetic correlation estimates appeared to have underestimation bias. In addition, the adjusted genetic correlation estimates between WC and EA, HC and EA,  WC and INC, and HC and INC, were significantly different from the unadjusted estimates with $p$-values of $0.006$, $0.021$, $0.028$, and $0.050$ (one-sided test).


\begin{table}[H]
\centering
\caption{\textbf{Adjusted heritability and genetic correlation with participation on 12 physical, sociodemographic, and lifestyle measures in the UKBB.} The block jackknife standard errors are shown in brackets. For binary traits, the
heritabilities of the liability scores are calculated.
The significant genetic correlations with participation before and after adjustment are highlighted in boldface. The phenotypes that are significantly correlated with participation after adjustment are highlighted in boldface.} 
\setlength{\tabcolsep}{1.5mm}
\resizebox{16.5cm}{!}{%
\begin{tabular}{cc|cc|cc|c}
\hline
\multirow{2}{*}{Phenotype} &	 \multirow{2}{*}{$\delta^\dagger$}	&\multicolumn{2}{c|}{$h^2$}&\multicolumn{2}{c|}{$\rho_{g}$}&	\multicolumn{1}{c}{$\rho_{e}$}\\
\cline{3-7}
&			&Original	& Adjusted & Original& Adjusted&	Adjusted\\
\hline
\textbf{BMI} & -0.138 & 0.251 (0.012) & 0.253 (0.012) & \textbf{-0.219} (0.073) & \textbf{-0.246} (0.066) & -0.030 (0.011)  \\
HGT & 0.237 & 0.544 (0.026) & 0.540 (0.026) & 0.024 (0.058) & 0.070 (0.055) & 0.155 (0.023)  \\
\textbf{WC} & -0.419 & 0.208 (0.010) & 0.216 (0.010) & \textbf{-0.244} (0.080) & \textbf{-0.355} (0.057) & -0.175 (0.011)  \\
\textbf{HC} & -0.336 & 0.231 (0.012) & 0.235 (0.012) & \textbf{-0.153 }(0.068) & \textbf{-0.245} (0.053) & -0.149 (0.011)  \\
\textbf{EA} & 0.438 & 0.203 (0.007) & 0.217 (0.008) & \textbf{0.366 }(0.096) & \textbf{0.467} (0.069) & 0.164 (0.012)  \\
\textbf{ES} & 0.089 & 0.021 (0.003) & 0.022 (0.002) & \textbf{0.352} (0.160) & \textbf{0.414} (0.134) & 0.024 (0.006)  \\
\textbf{INC} & 0.235 & 0.080 (0.004) & 0.086 (0.004) & \textbf{0.356} (0.109) & \textbf{0.443} (0.085) & 0.078 (0.009)  \\
\textbf{SMC} & -0.413 & 0.127 (0.008) & 0.131 (0.004) & -0.102 (0.086) & \textbf{-0.262} (0.083) & -0.192 (0.008)  \\
SMP & -0.043 & 0.117 (0.005) & 0.117 (0.003) & -0.011 (0.070) & -0.029 (0.066) & -0.020 (0.007)  \\
ALC & 0.131 & 0.080 (0.004) & 0.077 (0.004) &\textbf{ -0.249} (0.091) & -0.169 (0.104) & 0.090 (0.008)  \\
\textbf{WKP} & 0.073 & 0.077 (0.004) & 0.078 (0.003) & \textbf{0.201} (0.082) & \textbf{0.228} (0.074) & 0.015 (0.008)  \\
WKT & 0.093 & 0.041 (0.003) & 0.039 (0.003) & \textbf{-0.291} (0.119) & -0.211 (0.133) & 0.066 (0.007)  \\

\hline
\end{tabular}
}
 {\raggedright $^\dagger$ The mean shift between the phenotypes in the UKBB and HSE (UKBB minus HSE), which is calculated after rank-based inverse normal transformation,  covariates correction, and standardization with UKBB standard deviations. \par}
\label{tab:realdata}
\end{table}

\section{Discussion}
In the last decade, many methods have been proposed to estimate heritability and genetic correlation. Despite the success, estimates are typically derived under the assumption of random sampling, without taking PB into account. In addition, the non-random component underlying participation is often implicitly thought of as a function of environmental factors and established phenotypes such as EA, frequency of alcohol use, etc. Following this line of reasoning, to adjust for PB, a  recent 
study\citep{schoeler2022correction} applied IPW to the UKBB data, with a propensity score constructed from other phenotypes. Apart from the limitations that the propensity score was limited by the availability of phenotypes that could be harmonized between UKBB and other random sampling studies, most  importantly, the propensity score does not include genotypes. For this IPW adjustment to be sufficient for analyses that include genotypes, the main focus of genetic studies that include heritability and genetic correlation estimates, it requires that the genetic component underlying participation manifests its effect entirely through the propensity score. Under this assumption, the participation polygenic score constructed in Benonisdottir and Kong (2023) \cite{benonisdottir2022genetics} (denoted as pPGS), based on a GWAS that did not use any information on other phenotypes, should not have any predictive power on other phenotypes conditioning on the propensity score. 
We examined this with four variables that are significantly associated with the pPGS:
EA, BMI, and the invitation and participation in a physical activity study. 
When adjusting for the propensity score constructed by Schoeler et al (2023), the associations between the pPGS and the four variables shrunk by some degree, but remained highly 
statistically significant ($p$ ranges from $1.7\times 10^{-5}$ to $<2\times 10^{-16}$, see Supplementary Note 1.8 and Supplementary Table 2),
indicating that the assumption does not hold. Similar analyses were performed in Benonisdottir and Kong (2023) \citep{benonisdottir2022genetics} using EA in the place of the propensity score. These results support the belief that participation should be treated as a complex behavioral trait in its own right with its specific genetic component, and not simply a consequence of other established phenotypes.

In this article, we build a statistical model for understanding the influence on estimates of parameters involving phenotypes of interest. Conceptually, to properly adjust estimates of parameters that involve both genotypes and phenotypes, $e.g.$ heritability and genetic correlation, information on the differences between sample and population for both genotypes and phenotypes is necessary. For genotype differences, we utilize the IBD-based information on the genetic component of participation \citep{benonisdottir2022genetics}, which distinguishes our approach from existing methods such as IPW. For phenotype differences, we utilize the mean shifts from population to sample of the phenotypes of interest. Further research would be to explore what other information, if available, could be used to improve the adjustments and allows us to entertain more complex PB models. We applied our adjustment method 
to 12 UKBB phenotypes. We found 8 of the phenotypes were significantly associated with participation of the UKBB. We also found that, without adjustments, the heritability and the absolute value of the genetic correlation estimates had underestimation bias in the sample of participants.
{We note that LDSC estimates SNP-based heritability, $i.e.$ the proportion of phenotypic variance 
attributable to a given set of SNPs\citep{zhu2020statistical, song2022leveraging}. 
This would increase as the number of SNPs increases and is in general a fraction of the heritability.
Assuming the genetic architecture of the SNP set is representative of the entire genome, we believe
 the expected value of the ratio between the adjusted and unadjusted estimates should be stable as the number of SNPs increases. As to the adjusted genetic correlation estimates, they should be estimates for the full genetic components also.}



There are several limitations to the current approach. First, the IBD-based participation GWAS we used only captured direct effects, and we did not consider the indirect effects of participation. Second, the sample size of IBD siblings for deriving participation statistics is relatively limited. Although we have demonstrated via both simulations and real data applications that the standard errors before and after our adjustment are comparable, larger sample sizes will lead to more accurate results with smaller MSEs. Third, in order to adjust for PB, we leveraged the HSE dataset to derive the mean shift between UKBB and HSE. We assumed that the HSE dataset was based on random sampling, as it has incorporated weighting to account for nonresponse bias.
Violation of this assumption would reduce the effectiveness of the adjustments, but most likely they would still be in the right direction. Lastly, the adjustments were calculated assuming the invited list, about 9.5 million in size for the UKBB, is representative of the target population, about 21 million in size with the age constraint \citep{fry2017comparison}. Violation of this assumption does not affect the estimate of the genetic component underlying the overall PB that incorporates both invitation bias and agreeing to participate if invited bias. However, the estimate of the strength of this genetic component, $e.g.$ its heritability, could be impacted, and through that affect the other adjustments. The exact effect is mathematically complicated and depends on many factors, but we can get an idea by considering two extreme scenarios for the UKBB study. One extreme is that all those who were not invited would not have agreed to participate if invited anyway. In this case, the appropriate adjustments would be smaller than what are currently estimated, $e.g.$ the adjustment of heritability estimates for other phenotypes should be shrunk by about $25\%$. The other extreme is if all the bias came from the invitation list, and participation upon invitation was completely random. In that case, our current adjustments would be in the right direction but conservative, $i.e.$ smaller than what should be. Given that, we believe the adjustments we provided are reasonable. To do substantially better, more complicated modelling and additional information, such as phenotypic differences between target population and the invited list, are probably required. 

Despite the limitations, our model and method provide insights about  the genetic architecture underlying participation and other phenotypes. It is, we believe, an important first step towards adjusting PB in genetic studies in a way that does not rely entirely on phenotypic differences between sample and population. For future research, 
it is conceptually advantageous to develop statistical methods to investigate the underlying causality of participation and other phenotypes. In particular, it is often assumed that EA has a causal effect on participation \citep{schoeler2022correction}. By contrast,  whether and to what extent the inclination to participate, as a behavioral phenotype, has a causal impact on other phenotypes including EA has not been seriously considered. A deeper understanding of causality will be helpful in understanding the underlying mechanisms and dynamics of human behavior.

\section{Methods}

\subsection{UKBB summary statistics}
The GWAS summary statistics of UKBB participation were directly downloaded from GWAS catalog \citep{benonisdottir2022genetics}. Other UKBB summary statistics  based on the genetic data of 274,485 white British unrelated individuals in the UKBB after quality control \citep{bycroft2018uk}. The phenotypes of interest were adjusted for year of birth (data-field 34), age at recruitment (data-field 21022), and sex (data-field 31) when applicable. We further adjusted top 40 principal components \citep{bycroft2018uk}. Quantitative phenotypes except for EA, ALC, and WLP, were rank-based inverse normal transformed \citep{beasley2009rank} separately for each sex. A detailed description of the process of phenotypes is provided in Supplementary Note 1.6. We used the same quality filtering protocol for the sequence variants as Benonisdottir and Kong (2023) \citep{benonisdottir2022genetics}. The analysis was restricted to the set of 500,632 high-quality variants with the missing rates $<5\%$ and with minor allele frequencies $>1\%$. We used PLINK 1.90 to derive GWAS summary statistics \citep{chang2015second}.

\subsection{HSE datasets}
We obtained the anthropometric measures from 81,118 individuals from the Health Survey for England (HSE) for the years 2006-2010 \citep{hse2006, hse2007, hse2008, hse2009, hse2010}, which consists of an annual cross-sectional survey. The samples are a representative population of England through a two-stage random probability sampling process 
\citep{fry2017comparison}. The HSE data have incorporated weighting to account for nonresponse bias since 2003 \citep{national2013}. A detailed description of the collection of HSE datasets is in Supplementary Note 1.6. For the computation of the mean shift between HSE and UKBB, we restricted samples of white British ancestry and ages from 40 to 65, resulting in 20,208 individuals. The phenotypes in both HSE and UKBB were adjusted for sex, age, age$^2$, sex$*$age, and sex$*$age$^2$.

\subsection{LD score regression}
We used LD score regression (LDSC, v.1.0.1) to derive the estimates for heritability and genetic correlation, which ignores the effects of PB. In the simulations, the LD scores were computed with true  LD matrices.  The LD scores in real data analyses were computed by the Pan-UKB team \citep{panukbb} (downloaded on 7 April 2021).

\subsection{Polygenic score analysis}
We computed the pPGS with PLINK 1.90 \citep{chang2015second}, which summed over the weighted genotypes of the $500,632$ SNPs after quality control. The $z$ scores were used as weights, which were transformed from the combined $p$-values in the GWAS summary statistics of participation provided in Benonisdottir and Kong (2023) \citep{benonisdottir2022genetics}. The pPGS was standardized to have variance 1, and the relationship between the pPGS and EA, BMI, secondary invitation (binary), and secondary participation (binary) was estimated with a linear regression and logit regression in R (v.4.3.1), in the group of White British unrelateds. The sex, year of birth, age at recruitment, genotyping array, and 40 principal components were used as covariates.

\section{Data availability}
The  GWAS summary statistics for participation are available on the GWAS catalog under the accession codes GCST90267221 and GCST90267223. The individual-level UKBB data can be applied on their website (\url{http://www.ukbiobank.ac.uk/register-apply/}).

\section{Code availability}
We have developed an R package for adjusting the effects of PB on the estimation of heritability and genetic correlations, which is available at \url{https://github.com/shuangsong0110/ParticipationBias}.

\section{Acknowledgements}
This research has been conducted using the UK Biobank Resource (\url{www.ukbiobank.ac.uk}) under Application Number 68672.

\section{Competing Interests}
The authors declare that they have no competing interests.

\bibliographystyle{naturemag} 
\bibliography{template}

\begin{thebibliography}{10}
\expandafter\ifx\csname url\endcsname\relax
  \def\url#1{\texttt{#1}}\fi
\expandafter\ifx\csname urlprefix\endcsname\relax\def\urlprefix{URL }\fi
\providecommand{\bibinfo}[2]{#2}
\providecommand{\eprint}[2][]{\url{#2}}

\bibitem{sudlow2015uk}
\bibinfo{author}{Sudlow, C.} \emph{et~al.}
\newblock \bibinfo{title}{{UK} biobank: an open access resource for identifying the causes of a wide range of complex diseases of middle and old age}.
\newblock \emph{\bibinfo{journal}{PLoS Medicine}} \textbf{\bibinfo{volume}{12}}, \bibinfo{pages}{e1001779} (\bibinfo{year}{2015}).

\bibitem{bulik2015ld}
\bibinfo{author}{Bulik-Sullivan, B.~K.} \emph{et~al.}
\newblock \bibinfo{title}{{LD} score regression distinguishes confounding from polygenicity in genome-wide association studies}.
\newblock \emph{\bibinfo{journal}{Nature Genetics}} \textbf{\bibinfo{volume}{47}}, \bibinfo{pages}{291--295} (\bibinfo{year}{2015}).

\bibitem{kong2018nature}
\bibinfo{author}{Kong, A.} \emph{et~al.}
\newblock \bibinfo{title}{The nature of nurture: Effects of parental genotypes}.
\newblock \emph{\bibinfo{journal}{Science}} \textbf{\bibinfo{volume}{359}}, \bibinfo{pages}{424--428} (\bibinfo{year}{2018}).

\bibitem{abdellaoui2021dissecting}
\bibinfo{author}{Abdellaoui, A.} \& \bibinfo{author}{Verweij, K.~J.}
\newblock \bibinfo{title}{Dissecting polygenic signals from genome-wide association studies on human behaviour}.
\newblock \emph{\bibinfo{journal}{Nature Human Behaviour}} \textbf{\bibinfo{volume}{5}}, \bibinfo{pages}{686--694} (\bibinfo{year}{2021}).

\bibitem{border2022assortative}
\bibinfo{author}{Border, R.} \emph{et~al.}
\newblock \bibinfo{title}{Assortative mating biases marker-based heritability estimators}.
\newblock \emph{\bibinfo{journal}{Nature Communications}} \textbf{\bibinfo{volume}{13}}, \bibinfo{pages}{660} (\bibinfo{year}{2022}).

\bibitem{winship1992models}
\bibinfo{author}{Winship, C.} \& \bibinfo{author}{Mare, R.~D.}
\newblock \bibinfo{title}{Models for sample selection bias}.
\newblock \emph{\bibinfo{journal}{Annual Review of Sociology}} \textbf{\bibinfo{volume}{18}}, \bibinfo{pages}{327--350} (\bibinfo{year}{1992}).

\bibitem{swanson2012uk}
\bibinfo{author}{Swanson, J.~M.}
\newblock \bibinfo{title}{The {UK} {B}iobank and selection bias}.
\newblock \emph{\bibinfo{journal}{The Lancet}} \textbf{\bibinfo{volume}{380}}, \bibinfo{pages}{110} (\bibinfo{year}{2012}).

\bibitem{seaman2013review}
\bibinfo{author}{Seaman, S.~R.} \& \bibinfo{author}{White, I.~R.}
\newblock \bibinfo{title}{Review of inverse probability weighting for dealing with missing data}.
\newblock \emph{\bibinfo{journal}{Statistical Methods in Medical Research}} \textbf{\bibinfo{volume}{22}}, \bibinfo{pages}{278--295} (\bibinfo{year}{2013}).

\bibitem{van2022reweighting}
\bibinfo{author}{van Alten, S.}, \bibinfo{author}{Domingue, B.~W.}, \bibinfo{author}{Galama, T.~J.} \& \bibinfo{author}{Marees, A.~T.}
\newblock \bibinfo{title}{Reweighting the {UK} {B}iobank corrects for pervasive selection bias due to volunteering}.
\newblock \emph{\bibinfo{journal}{International Journal of Epidemiology}} \textbf{\bibinfo{volume}{53}}, \bibinfo{pages}{dyae054} (\bibinfo{year}{2024}).

\bibitem{bisgard1994mortality}
\bibinfo{author}{Bisgard, K.~M.}, \bibinfo{author}{Folsom, A.~R.}, \bibinfo{author}{Hong, C.-P.} \& \bibinfo{author}{Sellers, T.~A.}
\newblock \bibinfo{title}{Mortality and cancer rates in nonrespondents to a prospective study of older women: 5-year follow-up}.
\newblock \emph{\bibinfo{journal}{American Journal of Epidemiology}} \textbf{\bibinfo{volume}{139}}, \bibinfo{pages}{990--1000} (\bibinfo{year}{1994}).

\bibitem{manjer2001malmo}
\bibinfo{author}{Manjer, J.} \emph{et~al.}
\newblock \bibinfo{title}{The malm{\"o} diet and cancer study: representativity, cancer incidence and mortality in participants and non-participants}.
\newblock \emph{\bibinfo{journal}{European Journal of Cancer Prevention}} \bibinfo{pages}{489--499} (\bibinfo{year}{2001}).

\bibitem{drivsholm2006representativeness}
\bibinfo{author}{Drivsholm, T.} \emph{et~al.}
\newblock \bibinfo{title}{Representativeness in population-based studies: a detailed description of non-response in a {D}anish cohort study}.
\newblock \emph{\bibinfo{journal}{Scandinavian Journal of Public Health}} \textbf{\bibinfo{volume}{34}}, \bibinfo{pages}{623--631} (\bibinfo{year}{2006}).

\bibitem{fry2017comparison}
\bibinfo{author}{Fry, A.} \emph{et~al.}
\newblock \bibinfo{title}{Comparison of sociodemographic and health-related characteristics of uk biobank participants with those of the general population}.
\newblock \emph{\bibinfo{journal}{American Journal of Epidemiology}} \textbf{\bibinfo{volume}{186}}, \bibinfo{pages}{1026--1034} (\bibinfo{year}{2017}).

\bibitem{knudsen2010health}
\bibinfo{author}{Knudsen, A.~K.}, \bibinfo{author}{Hotopf, M.}, \bibinfo{author}{Skogen, J.~C.}, \bibinfo{author}{{\O}verland, S.} \& \bibinfo{author}{Mykletun, A.}
\newblock \bibinfo{title}{The health status of nonparticipants in a population-based health study: the {H}ordaland {H}ealth {S}tudy}.
\newblock \emph{\bibinfo{journal}{American Journal of Epidemiology}} \textbf{\bibinfo{volume}{172}}, \bibinfo{pages}{1306--1314} (\bibinfo{year}{2010}).

\bibitem{schoeler2022correction}
\bibinfo{author}{Schoeler, T.} \emph{et~al.}
\newblock \bibinfo{title}{Participation bias in the {UK B}iobank distorts genetic associations and downstream analyses}.
\newblock \emph{\bibinfo{journal}{Nature Human Behaviour}} \textbf{\bibinfo{volume}{7}}, \bibinfo{pages}{1216--1227} (\bibinfo{year}{2023}).

\bibitem{benonisdottir2022genetics}
\bibinfo{author}{Benonisdottir, S.} \& \bibinfo{author}{Kong, A.}
\newblock \bibinfo{title}{Studying the genetics of participation using footprints left on the ascertained genotypes}.
\newblock \emph{\bibinfo{journal}{Nature Genetics}} \textbf{\bibinfo{volume}{55}}, \bibinfo{pages}{1413--1420} (\bibinfo{year}{2023}).

\bibitem{birnbaum1950effect}
\bibinfo{author}{Birnbaum, Z.~W.}
\newblock \bibinfo{title}{Effect of linear truncation on a multinormal population}.
\newblock \emph{\bibinfo{journal}{The Annals of Mathematical Statistics}} \textbf{\bibinfo{volume}{21}}, \bibinfo{pages}{272--279} (\bibinfo{year}{1950}).

\bibitem{kan2017moments}
\bibinfo{author}{Kan, R.} \& \bibinfo{author}{Robotti, C.}
\newblock \bibinfo{title}{On moments of folded and truncated multivariate normal distributions}.
\newblock \emph{\bibinfo{journal}{Journal of Computational and Graphical Statistics}} \textbf{\bibinfo{volume}{26}}, \bibinfo{pages}{930--934} (\bibinfo{year}{2017}).

\bibitem{ni2018estimation}
\bibinfo{author}{Ni, G.} \emph{et~al.}
\newblock \bibinfo{title}{Estimation of genetic correlation via linkage disequilibrium score regression and genomic restricted maximum likelihood}.
\newblock \emph{\bibinfo{journal}{The American Journal of Human Genetics}} \textbf{\bibinfo{volume}{102}}, \bibinfo{pages}{1185--1194} (\bibinfo{year}{2018}).

\bibitem{zhu2020statistical}
\bibinfo{author}{Zhu, H.} \& \bibinfo{author}{Zhou, X.}
\newblock \bibinfo{title}{Statistical methods for {SNP} heritability estimation and partition: A review}.
\newblock \emph{\bibinfo{journal}{Computational and Structural Biotechnology Journal}} \textbf{\bibinfo{volume}{18}}, \bibinfo{pages}{1557--1568} (\bibinfo{year}{2020}).

\bibitem{song2022leveraging}
\bibinfo{author}{Song, S.}, \bibinfo{author}{Jiang, W.}, \bibinfo{author}{Zhang, Y.}, \bibinfo{author}{Hou, L.} \& \bibinfo{author}{Zhao, H.}
\newblock \bibinfo{title}{Leveraging {LD} eigenvalue regression to improve the estimation of {SNP} heritability and confounding inflation}.
\newblock \emph{\bibinfo{journal}{The American Journal of Human Genetics}} \textbf{\bibinfo{volume}{109}}, \bibinfo{pages}{802--811} (\bibinfo{year}{2022}).

\bibitem{bycroft2018uk}
\bibinfo{author}{Bycroft, C.} \emph{et~al.}
\newblock \bibinfo{title}{The {UK} {B}iobank resource with deep phenotyping and genomic data}.
\newblock \emph{\bibinfo{journal}{Nature}} \textbf{\bibinfo{volume}{562}}, \bibinfo{pages}{203--209} (\bibinfo{year}{2018}).

\bibitem{beasley2009rank}
\bibinfo{author}{Beasley, T.~M.}, \bibinfo{author}{Erickson, S.} \& \bibinfo{author}{Allison, D.~B.}
\newblock \bibinfo{title}{Rank-based inverse normal transformations are increasingly used, but are they merited?}
\newblock \emph{\bibinfo{journal}{Behavior Genetics}} \textbf{\bibinfo{volume}{39}}, \bibinfo{pages}{580--595} (\bibinfo{year}{2009}).

\bibitem{chang2015second}
\bibinfo{author}{Chang, C.~C.} \emph{et~al.}
\newblock \bibinfo{title}{Second-generation {PLINK}: rising to the challenge of larger and richer datasets}.
\newblock \emph{\bibinfo{journal}{Gigascience}} \textbf{\bibinfo{volume}{4}}, \bibinfo{pages}{s13742--015} (\bibinfo{year}{2015}).

\bibitem{hse2006}
\bibinfo{author}{{National Centre for Social Research}} \& \bibinfo{author}{{University College London, Department of Epidemiology and Public Health}}.
\newblock \bibinfo{title}{Health survey for {E}ngland, 2006}.
\newblock \bibinfo{howpublished}{[data collection]} (\bibinfo{year}{2011}).
\newblock \urlprefix\url{http://doi.org/10.5255/UKDA-SN-5809-1}.
\newblock \bibinfo{note}{SN: 5809}.

\bibitem{hse2007}
\bibinfo{author}{{National Centre for Social Research}} \& \bibinfo{author}{{University College London, Department of Epidemiology and Public Health}}.
\newblock \bibinfo{title}{Health survey for {E}ngland, 2007}.
\newblock \bibinfo{howpublished}{[data collection]} (\bibinfo{year}{2010}).
\newblock \urlprefix\url{http://doi.org/10.5255/UKDA-SN-6112-1}.
\newblock \bibinfo{note}{SN: 6112}.

\bibitem{hse2008}
\bibinfo{author}{{National Centre for Social Research}} \& \bibinfo{author}{{University College London, Department of Epidemiology and Public Health}}.
\newblock \bibinfo{title}{Health survey for {E}ngland, 2008}.
\newblock \bibinfo{howpublished}{[data collection]} (\bibinfo{year}{2013}).
\newblock \urlprefix\url{http://doi.org/10.5255/UKDA-SN-6397-2}.
\newblock \bibinfo{note}{SN: 6397}.

\bibitem{hse2009}
\bibinfo{author}{{National Centre for Social Research}} \& \bibinfo{author}{{University College London, Department of Epidemiology and Public Health}}.
\newblock \bibinfo{title}{Health survey for {E}ngland, 2009}.
\newblock \bibinfo{howpublished}{[data collection]} (\bibinfo{year}{2015}).
\newblock \urlprefix\url{http://doi.org/10.5255/UKDA-SN-6732-2}.
\newblock \bibinfo{note}{SN: 6732}.

\bibitem{hse2010}
\bibinfo{author}{{National Centre for Social Research}} \& \bibinfo{author}{{University College London, Department of Epidemiology and Public Health}}.
\newblock \bibinfo{title}{Health survey for {E}ngland, 2010}.
\newblock \bibinfo{howpublished}{[data collection]} (\bibinfo{year}{2015}).
\newblock \urlprefix\url{http://doi.org/10.5255/UKDA-SN-6986-3}.
\newblock \bibinfo{note}{SN: 6986}.

\bibitem{national2013}
\bibinfo{title}{National {C}entre for social research health survey for {E}ngland 2003. {V}olume 3. {M}ethodology and documentation {L}ondon, {U}nited {K}ingdom: Department of health; 2004.}
\newblock \bibinfo{howpublished}{\url{https://webarchive.nationalarchives.gov.uk/ukgwa/20121206162012/http://www.dh.gov.uk/prod_consum_dh/groups/dh_digitalassets/@dh/@en/documents/digitalasset/dh_4098912.pdf}}.
\newblock \bibinfo{note}{Published December 17, 2004. Accessed December 22, 2015.}

\bibitem{panukbb}
\bibinfo{title}{Pan-{UKB} team.}
\newblock \bibinfo{howpublished}{\url{https://pan.ukbb.broadinstitute.org.}}
\newblock \bibinfo{note}{2020.}

\end{thebibliography}

\end{document}